\documentclass[twocolumn,superscriptaddress,amsmath,amssymb,aps]{revtex4-2}

\usepackage[pdftex]{graphicx}
\usepackage{dcolumn}
\usepackage{bm}
\usepackage{float}
\usepackage{hyperref}
\usepackage{epstopdf}
\usepackage{multirow}
\usepackage{xr-hyper}
\newcommand{\ket}[1]{\left| #1 \right>}
\newcommand{\bra}[1]{\left< #1 \right|}

\newcommand{\Fig}[1]{Fig.\,\ref{#1}}
\newcommand{\Eq}[1]{Eq.\,\eqref{#1}}

\hypersetup{
            colorlinks=true,
            linkcolor=blue,
            anchorcolor=blue,
            citecolor=blue}
\begin{document}

\title{An intrinsic nonlinear Ohmic  current }
\author{YuanDong Wang}
\affiliation{School of Electronic, Electrical and Communication Engineering, University of Chinese Academy of Sciences, Beijing 100049, China}
\affiliation{School of Physical Sciences, University of Chinese Academy of Sciences, Beijing 100049, China}
\author{ZhiFan Zhang}
\affiliation{School of Physical Sciences, University of Chinese Academy of Sciences, Beijing 100049, China}
\author{Zhen-Gang Zhu}
\email{zgzhu@ucas.ac.cn}
\affiliation{School of Electronic, Electrical and Communication Engineering, University of Chinese Academy of Sciences, Beijing 100049, China}
\affiliation{School of Physical Sciences, University of Chinese Academy of Sciences, Beijing 100049, China}
\affiliation{CAS Center for Excellence in Topological Quantum Computation, University of Chinese Academy of Sciences, Beijing 100190, China}
\author{Gang Su}
\email{gsu@ucas.ac.cn}
\affiliation{School of Physical Sciences, University of Chinese Academy of Sciences, Beijing 100049, China}
\affiliation{CAS Center for Excellence in Topological Quantum Computation, University of Chinese Academy of Sciences, Beijing 100190, China}
\affiliation{Kavli Institute for Theoretical Sciences, University of Chinese Academy of Sciences, Beijing 100190, China}
\begin{abstract}
It is known that intrinsic currents in magnetic metals often appear in the direction perpendicular to the external field for linear and nonlinear responses. Distinct from three kinds of known nonlinear currents, namely, the Drude contribution, the Berry curvature dipole induced current and the Berry connection polarization induced current, here we report a  intrinsic nonlinear  current with breaking time-reversal symmetry. This new kind of intrinsic nonlinear current from the nontrivial Berry connection polarizability may emerge in the longitudinal or transverse direction, and both are dissipative Ohmic currents. We unveil 66 magnetic point group symmetries that can accommodate such nonlinear current, and possible candidate materials are proposed. This theory is also applied to observe the nonlinear current we proposed in one- and two-dimensional Dirac systems as examples.
\end{abstract}
\pacs{72.15.Qm,73.63.Kv,73.63.-b}
\maketitle

{\it{Introduction.}}
The anomalous Hall effect (AHE) is a fundamental transport phenomenon in which the charge current flows perpendicularly to the applied electric field without an external magnetic field, which is caused by the intrinsic or extrinsic mechanism, depending on whether the Hall conductivity is contributed by disorders \cite{RevModPhys.82.1539, RevModPhys.82.1959}. An essential consensus in last two decades is that the linear intrinsic AHE is related to the topology of energy bands. The semiclassical equations of motion describing the wave-packet dynamics are augmented by an anomalous velocity term that is perpendicular to the driving electric field \cite{PhysRevLett.75.1348, PhysRevB.53.7010, PhysRevB.59.14915}, thus the response current is of Hall-type. Recently, the nonlinear Hall responses arouse great interests. In particular, the nonlinear response can be the leading order at certain symmetries.
By now, three kinds of the 2nd-order DC current are discovered, say, the nonlinear Drude current which arises from  the intraband effects only, as a counterpart of the usual linear Drude current; the one from the Berry curvature dipole (BCD), which is a Hall-type current and proportional to the relaxation time \cite{PhysRevLett.115.216806, PhysRevLett.105.026805, ma2019observation, kang2019nonlinear}, and thus gives an extrinsic nonlinear Hall effect.
Such disorder-induced nonlinear Hall effect in time reversal symmetric systems was investigated in many studies  \cite{du2019disorder,PhysRevB.100.165422,isobe2020high, PhysRevLett.124.067203,PhysRevLett.124.067203, konig2021quantum,du2021nonlinear, du2021quantum}; and
the third one attributed to the Berry-connection polarizability (BCP) \cite{PhysRevLett.112.166601, PhysRevLett.127.277202, PhysRevLett.127.277201, PhysRevB.105.045118}, which is also a Hall-type current and independent of the relaxation time, to this end it is referred as the intrinsic nonlinear anomalous Hall effect (INAHE), recently it has been experimentally explored in \cite{wang2023quantum, gao2023quantum}.

For a long time, the \textit{Ohmic} response current, for which the conductivity tensor is symmetric in all indices \cite{10.21468/SciPostPhysCore.5.3.039}, is known as a dissipative term rather than a topological property. The \textit{Hall} response current, which is described by the antisymmetric part of the conductivity tensor, is commonly believed due to topological properties, such as those in quantum Hall effect, anomalous Hall effect, spin Hall effect, and nonlinear Hall effect, etc. Remarkably and surprisingly, in this work we found that there exists another intrinsic  nonlinear response current due to nontrivial BCP, which is of Ohmic with symmetric conductivity tensor.  Both longitudinal and transverse components give rise to \textit{dissipative} nonlinear currents from topological property.



The Ohmic nonlinear current identified in this work is intrinsic, which differs strikingly from the known  nonlinear Drude current \cite{ideue2017bulk,PhysRevB.104.054429,PhysRevX.11.011001} that is extrinsic.  Moreover, it differs remarkably from the known intrinsic nonlinear current like INAHE as well, because the INAHE is a genuine Hall effect and thus {\it{dissipationless}}.
We further analyze the symmetry properties of the intrinsic nonlinear Ohmic current (INOC) we proposed in 122 magnetic point groups (MPGs). More importantly, we point out that (besides the Drude term) the INOC exists uniquely in materials with the symmetry groups of $\bar{6}, 6^{\prime} /\text{m}, \bar{6}\text{m}2, \bar{6}\text{m}^{\prime} 2^{\prime}, 6^{\prime} /\text{m}\text{m}\text{m}^{\prime}, 23, \text{m}^{\prime} \bar{3}, 4^{\prime} 32^{\prime}, \bar{4}3\text{m}, \text{m}^{\prime} \bar{3}\text{m}$, in which the predicted INOC may be tested directly in experiments.


\begin{table*}[hptb]
	\renewcommand\arraystretch{2}
	\caption{List of constraints on the longitudinal and transverse component of the intrinsic nonlinear Ohmic current by the generators of magnetic point groups. The allowed (forbidden) conductivity tensors are indicated by $\checkmark$ ($\times$). }\label{mgp2}
	\begin{tabular*}{17cm}{@{\extracolsep{\fill}}p{0.8cm}ccccccccccccccccccc}
		\hline\hline
         & $\mathcal{P}$ & $C_2^y$ & $C_2^z$ & $\mathcal{P} C_2^y$ & $\mathcal{P} C_2^z$ & $C_3^z$ & $C_4^z$ & $\mathcal{P} C_4^z$ & $C_4^{z}\sigma_v$ & $\mathcal{T}$ & $\mathcal{PT}$& $C_2^y\mathcal{T}$& $C_2^z\mathcal{T}$ & $\mathcal{P} C_2^y\mathcal{T}$ & $\mathcal{P} C_2^z\mathcal{T}$ & $ C_3^z\mathcal{T}$ & $C_4^z\mathcal{T}$ & $\mathcal{P} C_4^z\mathcal{T}$& $C_4^z \sigma_v \mathcal{T}$
          \\
		\hline
        $\sigma_{\text{INOL}}^{xxx}$
        &$\times$ & $\times$ &$\times$ &$\checkmark$  &  $\checkmark$    & $\checkmark$ & $\times$ & $\times$& $\checkmark$&$\times$&$\checkmark$&$\checkmark$&$\checkmark$& $\times$& $\times$& $\times$& $\times$& $\times$& $\times$
        \\
        \hline
	  $\sigma_{\text{INOL}}^{yyy}$
	  & $\times$ &$\checkmark$  &$\times$ &$\times$ &  $\checkmark$  & $\checkmark$ & $\times$ & $\times$ &$\checkmark$&$\times$&$\checkmark$&$\times$&$\checkmark$&$\checkmark$& $\times$& $\times$& $\times$& $\times$& $\times$
	  \\
	  \hline
	  $\sigma_{\text{INOL}}^{xyy}$ & $\times$ & $\times$ & $\times$ & $\checkmark$ &  $\checkmark$  & $\checkmark$ & $\times$ & $\times$ &$\checkmark$&$\times$&$\checkmark$&$\checkmark$&$\checkmark$& $\times$& $\times$& $\times$& $\times$& $\times$& $\times$
	  \\
	  \hline
	  $\sigma_{\text{INOL}}^{yxx}$ & $\times$ & $\checkmark$ & $\times$ & $\times$ &  $\checkmark$  & $\checkmark$ & $\times$ & $\times$ &$\checkmark$&$\times$&$\checkmark$&$\times$&$\checkmark$&$\checkmark$& $\times$& $\times$& $\times$& $\times$& $\times$
 \\
		\hline\hline
	\end{tabular*}
\end{table*}

{\it{Intrinsic nonlinear Ohmic current.}}
The 2nd-order conductivity tensor is defined as the quadratic current response $\bm{J}$ to the electric field $\bm{E}$:
$
J^{(2),\alpha} = \sigma^{\alpha\alpha_1\alpha_2} E^{\alpha_1}E^{\alpha_2}
$  with $\alpha,\alpha_1,\alpha_2 \cdots \in \{x,y,z\}$  are the space indices.
Following the standard density matrix equations of motion approach \cite{PhysRevB.48.11705,PhysRevB.99.045121,PhysRevResearch.2.043081,PhysRevB.100.195117},  we present a detailed derivation of the 2nd-order conductivity tensor. In the static limit, we show that the 2nd-order conductivity tensor within isotropic relaxation approximation can be separated into the extrinsic and the intrinsic contributions as  $\sigma^{\alpha\alpha_1\alpha_2}=\sigma^{\alpha\alpha_1\alpha_2}_{\text{ext}}+\sigma^{\alpha\alpha_1\alpha_2}_{\text{int}}$. The extrinsic terms contain the Drude and the BCD-induced currents identified in previous studies \cite{PhysRevB.61.5337,PhysRevLett.123.246602,PhysRevResearch.2.043081,PhysRevX.11.011001} and  are reproduced in detail in the Supplemental Material \cite{supp}. The relaxation time $\tau$ dependence of the Drude is $\mathcal{O}(\tau^2)$, and $\mathcal{O}(\tau)$ for the BCD-induced currents. For the BCD-induced current, the $\mathcal{O}(\tau)$ dependence is related to the time reversal symmetry $\mathcal{T}$-even property.  Thus those terms dominate in clean metals.
The intrinsic part is composed of two terms given by
\begin{equation}\label{int}
\sigma^{\alpha\alpha_1\alpha_2}_{\text{int}}=\sigma^{\alpha\alpha_1\alpha_2}_{\text{INAHE}} + \sigma^{\alpha\alpha_1\alpha_2}_{\text{INOC}}.
\end{equation}
 In which the INAHE term is written as $\sigma^{\alpha\alpha_1\alpha_2}_{\rm{INAHE}}=\sum_{m}\frac{e^3}{\hbar}\int [d \bm{k} ]f_{  \bm{k}m}\Lambda^{\alpha\alpha_1 \alpha_2}_{\bm{k}m}$, with $\Lambda^{\alpha\alpha_1 \alpha_2}_{\bm{k}m}= f_{  \bm{k}m}(\partial^\alpha G_{\bm{k}m}^{\alpha_1\alpha_2}  -  \partial^{\alpha_1}G_{\bm{k}m}^{\alpha\alpha_2})$ and $G_{\bm{k}m}^{\alpha_1\alpha_2}=\sum_{n}2\text{Re(}\mathcal{A}_{  \bm{k}mn}^{\alpha_1}\mathcal{A}_{ \bm{k} nm}^{\alpha_2}/\varepsilon_{\bm{k}  mn})$ is the Berry connection polarizability \cite{PhysRevLett.112.166601, PhysRevLett.127.277202, PhysRevLett.127.277201, PhysRevB.105.045118}. The BCP is responsible to nonadiabatic Laudau-zener tunnelling processes \cite{kitamura2020nonreciprocal, PhysRevB.105.075201}.  $\varepsilon_{\bm{k}  mn}$ is defined as $\varepsilon_{\bm{k}  mn} = \varepsilon_{\bm{k}m}-\varepsilon_{\bm{k}n}.$ For convenience, we discard the $\bm{k}$ index in the following unless otherwise specified. 
The INAHE term is independent of $\tau$, and thus is intrinsic. Accordingly, $\sigma_{\text{INAHE}}$ requires $\mathcal{T}$-breaking and allows the $\mathcal{PT}$-symmetry.

The central result of this paper is the discovery of the Ohmic conductivity tensor in \Eq{int}, which is expressed as 
\begin{equation}
\sigma^{\alpha\alpha_1\alpha_2}_{\text{INOC}} = \sum_{m}\frac{e^3}{\hbar} \int [d \bm{k} ]\Gamma_{m}^{\alpha\alpha_1\alpha_2} ,
\end{equation}
with $f_{m}$ the equilibrium Fermi-Dirac distribution for the $m$-th energy band. The integrand $\Gamma_m$ is given by \cite{supp} (Noting that the same formula is independently obtained in \cite{lahiri2022intrinsic})
\begin{equation}\label{INOC-g}
\Gamma_{m}^{\alpha\alpha_1\alpha_2}=f_{  m}(\partial^{\alpha}G_{m}^{\alpha_1\alpha_2}+\partial^{\alpha_1}G_{m}^{\alpha\alpha_2}+\partial^{\alpha_2}G_{m}^{\alpha\alpha_1}),
\end{equation}
where $\mathcal{A}_{\bm{k}mn}^{\alpha}=i\langle u_{\bm{k}m}\ket{\partial^{\alpha} u_{\bm{k}n}}$ is the interband Berry connection with $\ket{u_{\bm{k}n}}$ the periodic part of the Bloch state (Latin indices $m$, $n, \cdots$ are used to label energy bands), and $\varepsilon_{mn}$ is the difference between energy bands $\varepsilon_{m}$ and $\varepsilon_{n}$.   One finds that by integration by part $\sigma_{\text{INOC}}$ is of the Fermi-surface effect. Noting that  an intrinsic longitudinal nonlinear current is reported \cite{holder2021mixed}, for which the cyclic permutation symmetry of spacial indices is violated that is identified as an  axial-gravitational anomaly \cite{PhysRevResearch.4.013217}. The semiclassical correspondence of such a current is the quantum geometric correction to the energy dispersion. However, for the INOC,   we have not find a semiclassical meaning of  INOC to the best of our knowledge \cite{gao2019semiclassical, 10.21468/SciPostPhys.14.4.082}.

It is seen that the INOC term is an intrinsic contribution in the sense that it is $O(\tau^0)$ and insensitive to the impurity scattering, manifesting as a parallel contribution to the INAHE.
According to \Eq{INOC-g}, we find that $\sigma^{\alpha\alpha_1\alpha_2}_{\text{INOC}}$ is  symmetric in all indices $\alpha$, $\alpha_1$, and $\alpha_2$, in this way it is recognized as the dissipative Ohmic current \cite{10.21468/SciPostPhysCore.5.3.039}. Noting that the direction of INOC is not confined to be along the longitudinal direction, but permitted in the transverse direction as well. However, the transverse component is not a genuine Hall effect as INAHE. For the latter the third-rank conductivity tensor is antisymmetric in two indices, then the current  is dissipationless and always perpendicular to electric field, irrespective of how the field is oriented relative to the crystal axes.   
One observes that  $\sigma_{\text{INOC}}$ is $\mathcal{T}$-odd \cite{supp}, and the combined symmetry of $\mathcal{T}$ and point group operation such as $\mathcal{PT}$-symmetry is allowed.

Owing to the $\mathcal{T}$-odd property, the symmetry transformations of $\sigma_{\text{INOC}}$ take the form of
\begin{equation}\label{INOC-st}
\sigma_{\text{INOC}}^{\alpha\beta\gamma}=\eta_{T}\sum_{\delta\nu\xi}D^{\alpha\delta}D^{\beta\nu}D^{\gamma\xi}\sigma_{\text{INOC}}^{\delta\nu\xi},
\end{equation}
where $D$ is the point group operation.  The $\mathcal{T}$-odd feature of $\sigma_{\text{INOC}}$ attributes to the factor $\eta_{T}$ defined by
\begin{equation}\label{ge}
\eta_{T}=\left\{
\begin{aligned}
1 & \quad {\text{primed operations}},  \\
-1 & \quad {\text{unprimed operations}}.
\end{aligned}
\right.
\end{equation}
By use of \Eq{ge}, the constrains of all 18 generators of MGPs on the in-plane $\sigma_{\text{INOC}}$ tensors are listed in Table \ref{mgp2}. Through Table \ref{mgp2}, the MPGs classified by the existence or non-existence of the four contributions of the 2nd-order conductivity tensors can be further deduced, as presented in Table II in Supplementary Material \cite{supp}.  Among all 122 MPGs, 66 of which allow $\sigma_{\text{INOC}}$. Particularly, the MPGs (second row of Table II in Supplementary Material), such as $\bar{3}^\prime \text{m}^\prime$, $\bar{6}$, $6^\prime /\text{m}$, allow $\sigma_{\text{INOC}}$ but forbid both $\sigma_{\text{BCD}}$ and $\sigma_{\text{INAHE}}$, leaving $\sigma_{\text{INOC}}$ as the unique  transverse 2nd-order response. In addition, some MPGs allow both INOC and INAHE contributions, enabling a comparison between these two intrinsic contributions.

\begin{figure}[tb]
\centering
\includegraphics [width=3.2in]{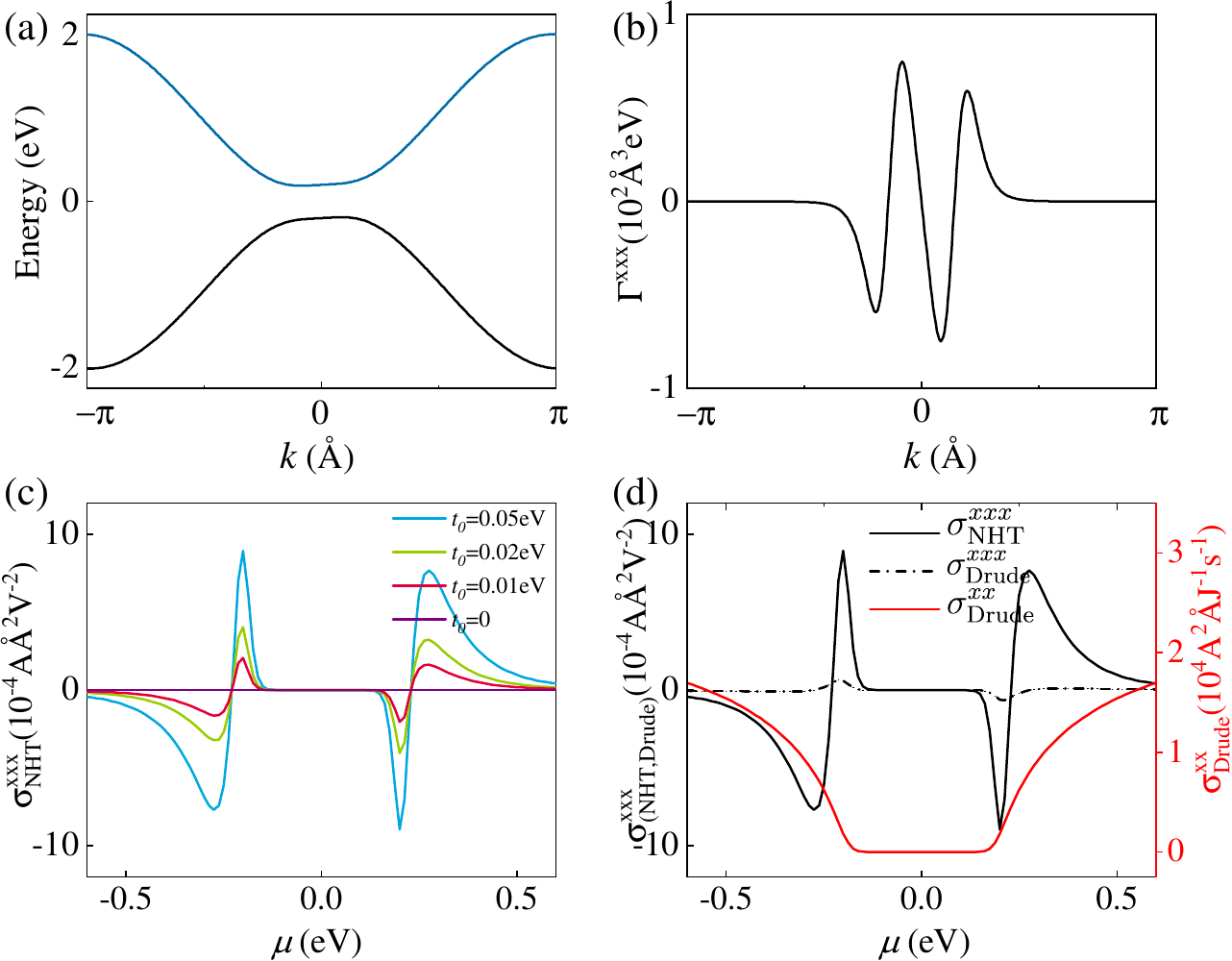}
\caption{(a) Band structure of the 1D toy model. (b) $\Gamma_{-}^{xxx}$ in the momentum space for the valence band. (c) The intrinsic longitudinal 2nd-order conductivity $\sigma_{\text{INOC}}^{xxx}$ versus the Fermi energy. (d) $\sigma_{\text{INOC}}^{xxx}$, $\sigma_{\text{Drude}}^{xxx}$ and the linear conductivity $\sigma_{\text{Drude}}^{xx}$ versus the Fermi energy.   The parameters  $t_0 = 0.05$ eV, $t_1 =t_2=2t_0$, $\Delta = 4t_0$ are used for Fig. 1(a),(b),(d), and  $t_1 =t_2=0.1$ eV, $\Delta = 0.2$ eV are used for Fig. 1(c).}\label{fig1}
\end{figure}

{\it{The intrinsic nonlinear Ohmic current in one-dimensional model.}}
A prominent feature of INOC is that it allows an intrinsic longitudinal response. To give a deeper understanding, we start with a one-dimensional (1D) toy model. The Hamiltonian is written as
\begin{equation}\label{1d}
H=t_0\sin{(ka)}  + [t_1 - t_2\cos{(ka)}]\tau_x  + \Delta\tau_z ,
\end{equation}
where $\tau$'s and $\sigma$'s are two sets of Pauli matrices. $t_0$, $t_1$, $t_2$ are parameters, $\Delta$ controls the gap, $a$  is the lattice constant, which is set to $a=1\rm{\AA}$. The energy dispersion is given as $\varepsilon_{\pm}(k)=t_{0}\sin{(ka)}\pm \sqrt{\Delta^2 + (t_1 - t_2 \cos{(ka)})^2}$, where $+ (-)$ sign corresponds to the conduction (valence) band. The first term $t_0\sin{(ka)}$ breaks the reflection symmetry and tilts the band structure, as manifested in \Fig{fig1}(a).

 Considering that the spin-freedom is not involved, the $\mathcal{PT}$-symmetry is represented by $\mathcal{PT}=\tau_x K$,  with $K$ the complex conjugate. One observes that $H$ respects the $\mathcal{PT}$ symmetry: $\mathcal{PT}H(k)(\mathcal{PT})^{-1}=H(k)$.
 The absence of reflection symmetry allows for a nonvanishing intrinsic longitudinal nonlinear conductivity  $\sigma^{xxx}_{\text{INOC}}$.

Combining the relations $\varepsilon_+(-k)=-\varepsilon_-(k)$ and $\mathcal{A}^{x}_{+}(-k)=-\mathcal{A}^{x}_{-}(k)$, it is easy to verify that the integrand $\Gamma^{xxx}_{\pm}$ is odd function with $\Gamma^{xxx}_{\pm}(k)=-\Gamma^{xxx}_{\pm}(-k)$, as illustrated in \Fig{fig1}(b). Therefore, it gives finite $\sigma^{xxx}_{\text{INOC}}$ when the reflection symmetry of dispersion is broken due to the tilt term. In \Fig{fig1}(c) we show $\sigma^{xxx}_{\text{INOC}}$ as a function of Fermi energy $\mu$ for different $t_0$. One can see that $\sigma^{xxx}_{\text{INOC}}$ is mostly concentrated around the gap, owing to its Fermi surface effect. The magnitude of $\sigma^{xxx}_{\text{INOC}}$ decreases as $t_0$ declines, as expected by symmetry analysis.

In this model,  the 2nd-order electric conductivities only contain $\sigma_{\text{INOC}}$ and Drude terms since both $\mathcal{P}$ and $\mathcal{T}$ symmetries are violated \cite{PhysRevB.99.045121,PhysRevResearch.2.043081}. The 2nd-order Drude conductivity is given as \cite{supp}
\begin{equation}
\sigma^{\alpha\alpha_1\alpha_2}_{\text{Drude}} = \frac{\tau^2e^3}{\hbar^3}\sum_m \int [d \bm{k} ]\varepsilon_{  m} \partial^{\alpha}\partial^{\alpha_1}\partial^{\alpha_2}f_{  m}.
\end{equation}
It is seen that the 2nd-order Drude conductivity arises only from the intraband effects.  The 2nd-order conductivity tensors $\sigma_{\text{INOC}}^{xxx}$, $\sigma_{\text{Drude}}^{xxx}$ are presented \Fig{fig1}(d), in which the relaxation time is set as $\tau = 10$ ps.
One observes for the 2nd-order Drude conductivity, there is  one peak and one valley around the gap, while for the INOC conductivity there are two peaks and two valleys around the gap.
This is because the interband Berry connections (for INOC conductivity) experience four times of drastic changes when the Fermi energy crosses the gap, which determines the behaviors of $\sigma_{\text{INOC}}^{xxx}$.
For $\sigma_{\text{Drude}}^{xxx}$, however, it is only determined by the band dispersion. The linear Drude conductivity $\sigma_{\text{Drude}}^{xx} = -\frac{e^2\tau}{\hbar^2}\int [d\bm{k}]\sum_m \varepsilon_{m}(\partial^{x}\partial^{x}f_{m})$ is also plotted in \Fig{fig1}(d). It is symmetric  about $\mu = 0$, as required by the second derivative on $f_{m}$. In practice, the intrinsic and extrinsic parts can be separated by their different scalings with the relaxation time. Since the linear (2nd-order) Drude conductivity is proportional to $\tau$ ($\tau^2$), the $\sigma_{\text{INOC}}$ is expected to dominate in low conducting samples (dirty limit).

\begin{figure}[tb]
\centering
\includegraphics [width=3.4in]{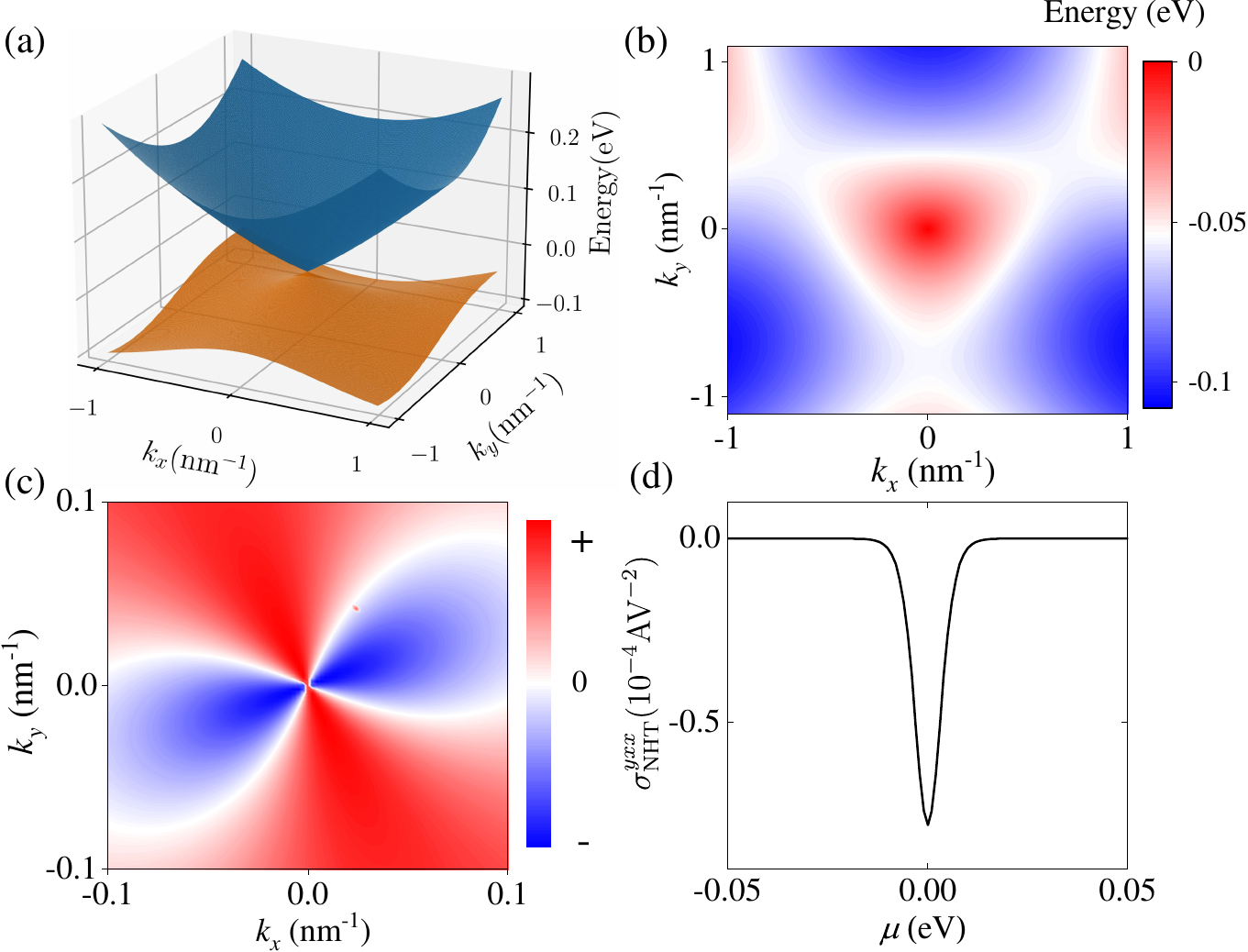}
\caption{(a) Band structure of the 2D four-band $k\cdot p$ model at the $k=0$ plane, where the upper two degenerate bands at $\bm{k}=0$ are plotted. (b) 2D contour plot of the energy dispersion for the valence band. (c) Band integrand $\Gamma^{xyy}$ for the valence band. (d) Calculated $\sigma_{\text{INOC}}^{xyy}$  as a function of Fermi energy. The parameters are: $e_1 = -1$ eV, $e_2 = 0.5e_1$, $e_3 =0$, $r_1 = r_2=r_3=r_4=-2e_1$. }\label{fig2}
\end{figure}

{\it{The transverse effect of intrinsic nonlinear Ohmic current in two-dimensional (2D) $k\cdot p$ model.}}
 It is intriguing to investigate the INOC in $\mathcal{T}$-breaking systems excluding the INAHE-induced current. As we demonstrated \cite{supp},  $C_3^z$-symmetry forbids $\sigma_{\text{INAHE}}$ but allows $\sigma_{\text{INOC}}$.
According to Table \ref{mgp2}, one of the MPGs that allows $\sigma_{\text{INOC}}$ but forbids $\sigma_{\text{INAHE}}$ and $\sigma_{\text{BCD}}$ is $\bar{3}^\prime \text{m}^\prime$.  One of the corresponding magnetic space group is $R$-$3^\prime c^\prime $(No.167.106). The $k\cdot p$ model of the magnetic space group $R$-$3^\prime c^\prime $ at $\Gamma$ point is found as \cite{PhysRevB.104.085137}
\begin{equation}
H(\bm{k}) =
\begin{pmatrix}
h_{11} & h_{12} & h_{13}  \\
h_{21} & h_{22} & h_{23} \\
h_{31} & h_{32} & h_{33}
\end{pmatrix}.
\end{equation}
The Hamiltonian is given  in form of the block matrices,
in which the blocks are written as $h_{11}=e_1$, $h_{22}=e_2$,

\begin{equation}
\begin{aligned}
h_{33} =& \frac{r_3}{2}\left(\frac{\sqrt{2}}{2}k_x \sigma_x + \sqrt{\frac{3}{2}}k_y \sigma_x - \sqrt{\frac{3}{2}}k_x \sigma_z + \frac{\sqrt{2}}{2} k_y \sigma_z \right)\\
&+e_3 \frac{\sqrt{2}}{2}\sigma_0,\\
h_{12} =& r_2 k_z, \\
h_{13} =&  \frac{r_3}{2}\left(-\sqrt{3}k_x M_{11} -k_x M_{12} +k_y M_{11}-\sqrt{3}k_y M_{12}\right),\\
h_{23} =&  \frac{r_4}{2}\left(k_x  M_{11} -\sqrt{3}k_x M_{12}+\sqrt{3}k_y   M_{11}+k_y M_{12}\right),
\end{aligned}
\end{equation}
where $\sigma$ are the Pauli matrixes, $M_{ij}$ are defined as the vectors of which only the $(i, j)$ entry is 1 and the rest are zero. $e_{1}$-$e_{3}$ and $r_{1}$-$r_4$ are the model parameters.
The model is two-fold degenerate at $\bm{k}=0$,  as shown in \Fig{fig2}(a). As a consequence, the Berry connections are mostly contributed by these two bands. In addition, the energy spectrum shows $C_{3}^z$-symmetry [as seen in \Fig{fig2}(b)], which is required by MPG $\bar{3}^\prime \text{m}^\prime$. In addition, the model contains $C_{2}^{x}$-symmetry, and the nonvanishing transverse response comes from $\sigma_{\text{INOC}}^{xyy}$ \cite{supp}.

The integrand $\Gamma^{xyy}$ in the momentum space for the lower band of these two bands is plotted in \Fig{fig2}(c). One observes that $\Gamma^{xyy}$ of this band does not preserve the inversion symmetry. As a consequence, the mismatch between the inversion antisymmetry of $\Gamma^{xyy}$ and the $C_{3}^{z}$ symmetry of $f_{m}$ results in a nonzero $\sigma_{\text{INOC}}^{xyy}$.
The $\sigma_{\text{INOC}}^{xyy}$ as a function of Fermi energy is shown in \Fig{fig2}(d). As the Fermi energy approaches to the degenerate point, the magnitude of the Berry connection becomes large due to the rapid changes of Bloch wave functions, leading to a large value of $\sigma_{\text{INOC}}^{xyy}$.

The material $\rm{AgRuO_{3}}$ reveals an antiferromagnetic transition at $T=342(3)$ K \cite{PhysRevB.103.214413}, for which the MPG is $\bar{3}^\prime \text{m}^\prime$ and it is possible to observe the INOC as the unique transverse response. We also expect the recently synthesized magnetoelectric  honeycomb $\rm{Mn_4 Ta_2 O_9}$ \cite{PhysRevB.98.134438} would be the similar platform to manifest the INOC. Other materials whose MPGs belong to the second row in Table II in Supplementary Material are also the candidate materials to observe the INOC.

\begin{figure}[tb]
\centering
\includegraphics [width=3.2in]{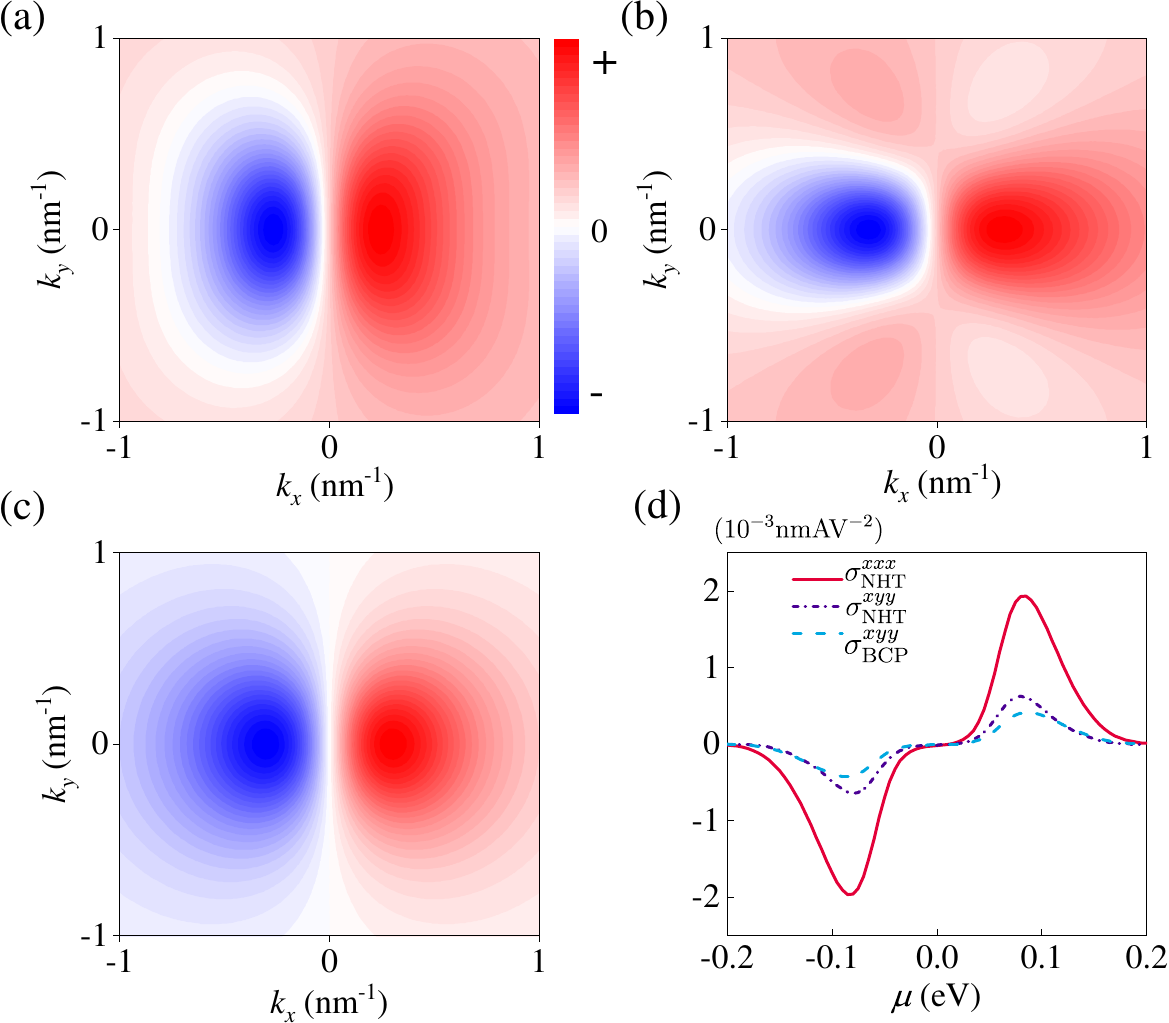}
\caption{(a)-(c) The lower band integrand $\Gamma^{xxx}$, $\Gamma^{xyy}$ and $\Lambda^{xyy}$ for $\sigma_{\text{INOC}}^{xxx}$, $\sigma_{\text{INOC}}^{xyy}$ and $\sigma_{\text{INAHE}}^{xyy}$. (d) Calculated $\sigma_{\text{INOC}}^{xxx}$, $\sigma_{\text{INOC}}^{xyy}$ and $\sigma_{INAHE}^{xyy}$ as a function of Fermi energy. The parameters are:  $v_x =v_y=1\times 10^{6}m/s$, $t=0.4v_x$, $\Delta = 0.06$ eV, $k_{B}T=0.01$ eV. }\label{fig3}
\end{figure}

{\it{Coexistence of the intrinsic nonlinear Ohmic current and the intrinsic nonlinear anomalous Hall effect.}}
Next we study two-dimensional systems where the intrinsic transverse 2nd-order responses induced by INAHE and INOC coexist. Here we consider a tilt four-band Dirac model, which describes orthorhombic \text{CuMnAs} \cite{tang2016dirac}. Such a model  is permitted a finite INAHE \cite{PhysRevLett.127.277202}. The Hamiltonian is given by
\begin{equation}
H(\bm{k}) = tk_x + v_x k_x \tau_x + v_y k_y \tau_y \sigma_x + \Delta \tau_z,
\end{equation}
where $(k_x, k_y)$ are the wave vectors, $\tau$ and $\sigma$ are two sets of Pauli matrices, $t$, $v_x$, $v_y$ and $\Delta$ are the model parameters.    $t$ tilts the system  along the $x$ direction, and $\Delta$ controls the gap. The dispersion is given by $E_{\pm}(\bm{k})  =tk_x \pm \sqrt{v_{x}^2 k_{x}^2 +v_{y}^2 k_{y}^2 +\Delta^2}$. Due to the presence of the spin freedom, the $\mathcal{PT}$ operation is represented by   $-i\sigma_y K$. By imposing the $\mathcal{PT}$ operation, one can verify that $\mathcal{PT}H(\bm{k})(\mathcal{PT})^{-1}=H(\bm{k})$, satisfying the $\mathcal{PT}$ symmetry. As a consequence, with each band two fold degeneracies appear at each momentum $\bm{k}$ due to the Kramers degeneracy.
In addition, the model holds $\sigma_{y}$ symmetry, therefore both $\sigma_{\text{INOC}}^{yxx}$ and $\sigma_{\text{INAHE}}^{yxx}$ are forced to vanish and the only nonvanishing transverse response comes from $\sigma_{\text{INOC}}^{xyy}$ and $\sigma_{\text{INAHE}}^{xyy}$ \cite{supp, PhysRevLett.127.277202}. For the longitudinal response, $\sigma_{\text{INOC}}^{xxx}$ is the only nonvanishing contribution for similar considerations.

For $\sigma_{\text{INOC}}^{xxx}$ and $\sigma_{\text{INOC}}^{xyy}$, the integrand $\Gamma^{xxx}$ for the lower band as well as $\Gamma^{xyy}$  are plotted in \Fig{fig3}(a) and \Fig{fig3}(b). Both show a dipole pattern, and the dipoles are mostly concentrated around the gap region.
Notably, one observes that both $\Gamma^{xxx}$ and $\Gamma^{xyy}$ are odd functions of $k_x$, which are similar to $\Lambda^{xyy}$, as shown in \Fig{fig3}(c).
Nevertheless, $\Gamma^{xxx}$, $\Gamma^{xyy}$ and $\Lambda^{xyy}$ are antisymmetric in the presence of inversion symmetry in momentum space. In the absence of the tilting, the corresponding currents vanishes when the integral is taking over the momentum. However, there should be finite corresponding currents due to the broken inversion symmetry when tilting term is introduced. As the Fermi energy changes, $\sigma_{\text{INOC}}^{xxx}$, $\sigma_{\text{INOC}}^{xyy}$ and $\sigma_{\text{INAHE}}^{xyy}$ show a similar Fermi-surface-dependent feature, as plotted in \Fig{fig3}(d).

{\it{Discussions.}}
%
%
The intrinsic nonlinear Ohmic current should be present in a large class of three-dimensional materials with time-reversal symmetry breaking. This present study illuminates a deeper insight on the intrinsic currents in ferromagnetic and antiferromagnetic materials. To find a semiclassical correspondence of the INOC is of particular interest. An additional natural extension of our work is the inclusion of spin effect. Recently, the nonlinear spin effects in $\mathcal{T}$-breaking system are attracting much attention, such as the nonlinear spin current induced by BCD in antiferromagnetic metals \cite{PhysRevB.106.024405} and nonlinear spin magnetoelectric effect induced by BCP \cite{xiao2022intrinsic}. The derivation of the spin version of the INOC might be straightforward. Likewise, a temperature gradient can also be capable in the generation of INOC. Based on the recently developed nonlinear thermal transport theory \cite{PhysRevB.101.155204,PhysRevB.106.035148}, the thermal counterpart of the INOC deserves a future study.

This work is supported in part by the NSFC (Grant Nos. 11974348, 11674317, and 11834014), and the National Key R$\&$D Program of China (Grant No. 2018FYA0305800). It is also supported by the Fundamental Research Funds for the Central Universities, and the Strategic Priority Research Program of CAS (Grant Nos. XDB28000000, and and XDB33000000).

%

\newpage
\begin{widetext}
\begin{center}
\textbf{\large  ONLINE SUPPLEMENTARY MATERIAL FOR
  \\[.2cm] Supplementary Material for "An  Intrinsic nonlinear Ohmic Current"}
\\[0.2cm]
YuanDong Wang,$^{1,2}$, ZhiFan Zhang$^{2}$, Zhen-Gang Zhu$^{1,2,3,*}$ and Gang Su$^{2,3,4,\dagger}$
\\[0.2cm]
{\small \it $^{1}$ School of Electronic, Electrical and Communication Engineering, University of Chinese Academy of Sciences, Beijing 100049, China}

{\small \it $^{2}$ School of Physical Sciences, University of Chinese Academy of Sciences, Beijing 100049, China}

{\small \it $^{3}$ CAS Center for Excellence in Topological Quantum Computation, University of Chinese Academy of Sciences, Beijing 100190, China}

{\small \it $^{4}$ Kavli Institute for Theoretical Sciences, University of Chinese Academy of Sciences, Beijing 100190, China}

\end{center}
\end{widetext}

\setcounter{equation}{0}
\setcounter{figure}{0}
\setcounter{table}{0}
\setcounter{page}{1}
\makeatletter
\renewcommand{\theequation}{S\arabic{equation}}
\renewcommand{\thefigure}{S\arabic{figure}}
\begin{widetext}

\tableofcontents

\section{Derivation of the second-order conductivity in the DC limit}\label{apd1}
\subsection{Reduced density matrix equations of motion}
By use of the reduced density matrix equations of motion,
we reproduce the expression for the 2nd-order electric conductivity \cite{PhysRevB.48.11705,PhysRevB.99.045121,PhysRevResearch.2.043081,PhysRevB.100.195117}.
We write down Hamiltonian in momentum space
\begin{equation}
\hat{H}_{0}=\sum_{m}\int [d\bm{k}]\varepsilon_{\bm{k}m}c_{\bm{k}m}^\dagger c_{\bm{k}m},
\end{equation}
with $\varepsilon_{\bm{k}m}$ the dispersion for $m$ band.  There are two standard ways to introduce the perturbation of the electric field within the framework of independent electrons. The first is so-called the velocity gauge with the momentum  substituted by
\begin{equation}
\hat{H}=\hat{H}_{0}(\bm{k}-e\bm{A}(t)),
\end{equation}
where $\bm{A}(t)$ is the magnetic vector potential. The other is the length gauge, where the electric field $\bm{E}(t)$ is coupled to the position operator $\hat{\bm{r}}$:
\begin{equation}\label{lenh}
\hat{H}=\hat{H}_{0}+ie\bm{E}(t)\cdot\hat{\bm{r}},
\end{equation}
which is known  as the dipole Hamiltonian.
The two descriptions of Hamiltonian can be shown equal by use of a gauge transformation \cite{PhysRevB.96.035431,PhysRevB.97.235446},  therefore the choice of the two gauges should give identical results.

In the following we adopt the length gauge. An obvious problem is that the matrix element of the position operator $\hat{\bm{r}}$ is ill defined. To this end, the covariant
derivative operator is introduced to overcome this difficulty \cite{PhysRevB.91.235320,PhysRevB.96.035431,PhysRevB.99.045121}.
 In the Bloch representation, it is defined as
\begin{equation}
\hat{\bm{D}}_{\bm{k}mn}=\delta_{mn}\nabla_{\bm{k}}-i\bm{\mathcal{A}}_{\bm{k}mn},
\end{equation}
where $\bm{\mathcal{A}}_{\bm{k}}$ is the Berry connection whose matrix elements are $\bm{\mathcal{A}}_{\bm{k}mn}=i\langle u_{\bm{k}m}\ket{\bm{\nabla}_{\bm{k}}u_{\bm{k}n}}$, with $\ket{u_{\bm{k}n}}$ the periodic part of the Bloch functions.
The covariant derivative operator acts on operators via
\begin{equation}\label{cov-de}
\begin{aligned}
[\hat{\bm{D}}_{\bm{k}},\hat{\mathcal{O}}(\bm{k})]_{mn}=\nabla_{\bm{k}}\mathcal{O}(\bm{k})_{mn}-i[\bm{\mathcal{A}}_{\bm{k}},\hat{\mathcal{O}}(\bm{k})]_{mn}.
\end{aligned}
\end{equation}
We adopt the reduced density matrix (RDM) equations of motion approach to calculate the nonlinear thermal response in length gauge. The RDM in band space is given by the average of the product of a creation and a destruction operator in Bloch states
\begin{equation}
\rho_{\bm{k}mn}(t)\equiv \langle c_{m\bm{k}}^{\dagger}(t)c_{n\bm{k}}(t) \rangle.
\end{equation}
With the definition of RDM, the expectation value of an operator $\mathcal{O}$ can be calculated as
\begin{equation}
\langle \hat{\mathcal{O}} \rangle = \text{tr}[\hat{\rho} \hat{\mathcal{O}}],
\end{equation}
where $\hat{\rho}$, the full many-body density matrix, can be written as
\begin{equation}
\hat{\rho} = \sum_{n}p_n \ket{\psi_n}\bra{\psi_n},
\end{equation}
with $\{\ket{\psi_n}\}$ a set of complete basis.
In the Schr{\"o}dinger picture, the time evolution of $\hat{\rho}$ is determined by the dynamics of $\ket{\psi_n}$:
\begin{equation}
i\hbar\frac{\partial}{\partial t}\ket{\psi_n (t)}=\hat{H}(t)\ket{\psi_n (t)}.
\end{equation}
The velocity operator is defined through
\begin{equation}
\hat{v}^{\alpha}=\partial_t \hat{r}^\alpha= \frac{1}{i\hbar}[\hat{r}^\alpha,\hat{H}].
\end{equation}
By use of the covariant derivative operator \Eq{cov-de}, its matrix elements  turn out to be
\begin{equation}\label{velo1}
\hat{v}^{\alpha}_{mn}=\frac{1}{\hbar}[\hat{D}^{\alpha},\hat{H}_{0}]_{mn}=\frac{1}{\hbar}\left[\delta_{mn}\partial^{\alpha}\varepsilon_{\bm{k}m}-i\mathcal{A}_{\bm{k}mn}^{\alpha}\varepsilon_{\bm{k}nm}\right].
\end{equation}
The electric current can be obtained from the RDMs and velocity operator as traces over band space:
\begin{equation}\label{current}
J^\alpha (t)=-e\int [d\bm{k}]\text{Tr}[\hat{\rho}(t)\hat{v}^\alpha].
\end{equation}
The equation of motion of the full many-body density matrix takes the form
\begin{equation}\label{rdms}
i\hbar\frac{\partial \hat{\rho}(t)}{\partial t}=[\hat{H}_{0},\hat{\rho}(t)]+ieE^\alpha [\hat{D}^\alpha ,\hat{\rho}(t)].
\end{equation}
Where  the summation of repeated space indices is indicated.
After computing the commutator on the r.h.s of \Eq{rdms}, we arrive at the closed equations of motion for the RDM:
\begin{equation}
\left[\frac{\partial}{\partial t}+i\frac{1}{\hbar}\varepsilon_{\bm{k}mn}\right]\rho_{\bm{k}mn}(t)= \frac{e}{\hbar}E^{\alpha} [\hat{D}^\alpha ,\rho(t)]_{\bm{k}mn}.
\label{iterativeRhon}
\end{equation}
Here, we define $\varepsilon_{\bm{k}mn}= \varepsilon_{\bm{k}m}-\varepsilon_{\bm{k}n}$.
  For convenience,   we discard the $\bm{k}$ index in the following unless otherwise specified.  By use of the Fourier transformation
\begin{equation}
\rho_{mn}(\omega)=\int\frac{d\omega}{2\pi}e^{-i\omega t}\rho_{mn}(t),
\end{equation}
the equations of motion in the frequency domain is
\begin{equation}
\rho_{mn}(\omega)=-\frac{e}{\hbar\omega - \varepsilon_{mn}}\int \frac{d\omega_1}{2\pi}E^{\alpha_1}(\omega_1)[\hat{D}^{\alpha_1} , \hat{\rho}(\omega-\omega_1)]_{mn}.
\end{equation}
  The density matrix is expanded into contributions of different powers on the external field as $\rho = \sum_{i=0}\rho^{(i)}$, where the zero-order one  is the the Fermi-Dirac distribution function times the unit matrix in band space $\rho^{(0)}_{\bm{k}mn}=f_{\bm{k}m}\delta_{mn}$. The recursion equations for the RDM are obtained as
\begin{equation}
\rho^{(n)}_{mn}(\omega)=-ed_{mn}(\omega)\int \frac{d\omega_1}{2\pi}E^{\alpha_1}(\omega_1)[\hat{D}^{\alpha_1} , \hat{\rho}^{(n-1)}(\omega-\omega_1)]_{mn}.
\end{equation}
with the matrix $d_{mn}(\omega)$ defined through
\begin{equation}
d_{mn}(\omega)=\frac{1}{\hbar\omega-\varepsilon_{mn}}.
\end{equation}
With the definition \Eq{current},  the current in terms of the Fourier components is written as
\begin{equation}
\begin{aligned}
J^{\alpha}(\omega)=\int d\omega_1  \sigma^{\alpha\alpha_1}(\omega;\omega_1) E^{\alpha_1}(\omega_1)+\int d\omega_1  d\omega_2 \sigma^{\alpha\alpha_1\alpha_2}(\omega;\omega_1,\omega_2) E^{\alpha_1}(\omega_1)E^{\alpha_2}(\omega_2)+\cdots ,
\end{aligned}
\end{equation}
in which the conductivity tensors are defined as the coefficients in an expansion of the current in powers of the electric field.

\subsection{Linear response}

The linear current is obtained in form of the 1st-order RDMs as
\begin{equation}\label{je-1}
J^{(1),\alpha}(\omega)=\int [d\bm{k}]\text{Tr}[\rho^{(1)}(\omega)(-e\hat{v}^{\alpha})],
\end{equation}
with response coefficients determined by the Kubo formula.
When we consider the static limit, from \label{rdm} the first order RDM  is found as
\begin{equation}\label{rho1}
\rho^{(1)}_{mn}=ie E^{\alpha_1}\cdot\left(\delta_{mn}\frac{1}{\hbar \omega}\partial^{\alpha_1}f_{m}-i \frac{1}{\hbar\omega -   \varepsilon_{mn}}\mathcal{A}^{\alpha_1}_{mn}f_{nm}\right).
\end{equation}
Here, we define $f_{nm}=f_{n}-f_{m}$.
Combining \Eq{je-1}, \Eq{velo1} and \Eq{rho1},  one observe that the linear current is contributed by two parts:
\begin{equation}
J^{(1),\alpha} =E^{\alpha_1}(\omega) [\chi_{1}^{\alpha\alpha_1}(\omega) + \chi_{  2}^{\alpha\alpha_1}(\omega)].
\end{equation}
Where the first term  is written as
\begin{equation}
\chi_{ 1}^{\alpha\alpha_1}(\omega) = -i\frac{e^2}{\hbar^2}\int [d\bm{k}]\frac{1}{\omega}\sum_m (\partial^{\alpha}\varepsilon_{m})(\partial^{\alpha_1}f_{m}).
\end{equation}
And the second term is related to the Berry connection
\begin{equation}
\chi_{ 2}^{\alpha\alpha_1}(\omega)  = i\frac{e^2}{\hbar} \sum_{m}\int [d\bm{k}]\frac{\varepsilon_{mn}}{\hbar\omega -\varepsilon_{mn}}\mathcal{A}^{\alpha_1}_{mn}\mathcal{A}^{\alpha}_{nm}f_{mn}.
\end{equation}

Now we consider the static limit, where the scattering rate is  introduced by replacing the matrix $d_{mn}$ as
\begin{equation}
d_{mn}(\omega)=\frac{1}{\hbar\omega+i\gamma-\varepsilon_{mn}}.
\end{equation}
In which $\gamma$  represents the relaxation rate, and the relaxation time can be approximately evaluated as $\tau=1/\gamma$. In our paper we focus on the simple case that the scattering of the impurities is phenomenologically integrated into $\gamma$. Specifically, when the scattering potential is explicitly written in Hamiltonian, the scattering of impurities can be formally treated in the equations of motion, as demonstrated in Ref.\cite{PhysRevB.96.235134, PhysRevB.100.195117, PhysRevResearch.2.033100}. In the static limit, one observes that the Drude contribution $\chi_{1}^{\alpha\alpha_1}$ is linearly dependent on $\tau$.
$\chi_{ 2}^{\alpha\alpha_1}(\omega)$  becomes
\begin{equation}
\chi_{ 2}^{\alpha\alpha_1}\xrightarrow[\text{DC}]{} \frac{e^2}{\hbar} \sum_{m}\int [d\bm{k}]\Omega^{\gamma}_{m}f_m,
\end{equation}
where the Berry curvature is defined as $\Omega^{\gamma}_{m}=\epsilon^{\alpha\alpha_1\gamma}\partial^{\alpha}\mathcal{A}^{\alpha_1 }_{mm}$, with  $\epsilon^{\alpha\alpha_1\gamma}$ the Levi-Civita symbol. It is seen that the intrinsic term in the static limit is independent of $\tau$, resulting the usual linear anomalous Hall effect.

\subsection{Second-order response}
Now we turn to the second-order response and demonstrate the emergence of BCD-induced current, INAHE and the newly discovered INOC under different symmetry.
The second-order current is written as
\begin{equation}\label{je-2}
J^{(2),\alpha}=\int [d\bm{k}]\text{Tr}[\rho^{(2)}_{  }(-e\hat{v}^{\alpha})].
\end{equation}
By use of the recursive relation \Eq{rdms}, the 2nd-order RDM is contributed by four terms:
\begin{equation}
\begin{aligned}
\rho^{(2)}_{  }(\omega)=e^2 \int \frac{d\omega_1}{2\pi} \int \frac{d\omega_2}{2\pi} E^{\alpha_1}(\omega_1)E^{ \alpha_2}(\omega_2)[\varrho^{1,\alpha_1\alpha_2}_{  }(\omega_1,\omega_2)+\varrho^{2,\alpha_1\alpha_2}_{  }(\omega_1,\omega_2)+\varrho^{3,\alpha_1\alpha_2}_{  }(\omega_1,\omega_2)+\varrho^{4,\alpha_1\alpha_2}_{  }(\omega_1,\omega_2)].
\end{aligned}
\end{equation}
In which the energy conservation $\omega = \omega_1 + \omega_2 $ is indicated.
The expressions for each term are obtained as
\begin{equation}
\begin{aligned}
\varrho^{1,\alpha_1\alpha_2}_{  mn}(\omega_1,\omega_2)=& -d_{mn}(\omega)d_{mn}(\omega_2) (\partial^{\alpha_1}\partial^{\alpha_2}f_{  m})\delta_{mn},\\
\varrho^{2,\alpha_1\alpha_2}_{  mn}(\omega_1,\omega_2)=& id_{mn}(\omega)d_{mm}(\omega_2)\mathcal{A}_{  mn}^{\alpha_1}(\partial^{\alpha_2}f_{  nm}),\\
\varrho^{3,\alpha_1\alpha_2}_{  mn}(\omega_1,\omega_2)=& id_{mn}(\omega)\partial^{\alpha_1}\left[d_{mn}(\omega_2)  \mathcal{A}_{  mn}^{\alpha_2}f_{  nm}\right],\\
\varrho^{4,\alpha_1\alpha_2}_{  mn}(\omega_1,\omega_2)=& \sum_{p} d_{mn}(\omega)\left[\mathcal{A}_{  mp}^{\alpha_1}d_{pn}(\omega_2)\mathcal{A}_{  pn}^{\alpha_2}f_{  np} -\mathcal{A}_{  pn}^{\alpha_1}d_{mp}(\omega_2)\mathcal{A}_{  mp}^{\alpha_2}f_{  pm}\right].
\end{aligned}
\end{equation}
Without written explicitly, the symmetrize procedure $[\alpha_1, \omega_1 ] \leftrightarrow [\alpha_2,\omega_2]$ is indicated.
In $\varrho^{1,\alpha_1\alpha_2}_{  mn}$ only the diagonal components of the covariant derivative operator are involved, as a result it is of two $k$ derivatives. For $\varrho^{2,\alpha_1\alpha_2}_{  mn}$ and $\varrho^{3,\alpha_1\alpha_2}_{  mn}$, one diagonal and one non-diagonal term are combined. While for $\varrho^{4,\alpha_1\alpha_2}_{  mn}$ no diagonal component is involved and it gathers all the interband (Berry connections) factors.
 Together with \Eq{velo1} and \Eq{je-2}, the second-order electric  current is given as
\begin{equation}
J^{(2),\alpha}(\omega_1,\omega_2) = E^{\alpha_1}(\omega_1)E^{\alpha_2}(\omega_2)[\chi_{1}^{\alpha\alpha_1\alpha_2}(\omega_1,\omega_2) + \chi_{2}^{\alpha\alpha_1\alpha_2}(\omega_1,\omega_2)+\chi_{3}^{\alpha\alpha_1\alpha_2}(\omega_1,\omega_2) + \chi_{4}^{\alpha\alpha_1\alpha_2}(\omega_1,\omega_2)+ \chi_{5}^{\alpha\alpha_1\alpha_2}(\omega_1,\omega_2)],
\end{equation}
with
\begin{eqnarray}
\chi^{\alpha\alpha_1\alpha_2}_{1}(\omega_1,\omega_2)&=& \sum_m \frac{e^3}{\hbar} \int [d \bm{k} ] d_{mn}(\omega)d_{mn}(\omega_2) (\partial^{\alpha}\varepsilon_{  m})  (\partial^{\alpha_1}\partial^{\alpha_2}f_{  m})\delta_{mn},\\
\chi^{\alpha\alpha_1\alpha_2}_{2}(\omega_1,\omega_2)&=& \sum_{m,n}\frac{e^3}{\hbar} \int [d \bm{k} ]-d_{mn}(\omega)d_{mm}(\omega_2)\varepsilon_{  mn}\mathcal{A}_{  nm}^{\alpha}\mathcal{A}_{  mn}^{\alpha_1}\partial^{\alpha_2}f_{  nm},\\
\chi^{\alpha\alpha_1\alpha_2}_{3}(\omega_1,\omega_2)&=& \sum_{m,n}\frac{e^3}{\hbar}\int [d \bm{k} ]-d_{mn}(\omega)\varepsilon_{  mn}\partial^{\alpha_1}\left[d_{mn}(\omega_2)  \mathcal{A}_{  mn}^{\alpha_2}f_{  nm}\right]\mathcal{A}_{  nm}^{\alpha},\\
\chi^{\alpha\alpha_1\alpha_2}_{4}(\omega_1,\omega_2)&=& \sum_{m,n,p} i\frac{e^3}{\hbar}\int [d \bm{k} ] d_{mn}(\omega)\varepsilon_{  mn}\mathcal{A}_{  nm}^{\alpha}\left[\mathcal{A}_{  mp}^{\alpha_1}d_{pn}(\omega_2)\mathcal{A}_{  pn}^{\alpha_2}f_{  np}-\mathcal{A}_{  pn}^{\alpha_1}d_{mp}(\omega_2)\mathcal{A}_{  mp}^{\alpha_2}f_{  pm}\right]\\
\chi^{\alpha\alpha_1\alpha_2}_{5}(\omega_1,\omega_2)&=& \sum_{m,n}\frac{e^3}{\hbar}\int [d \bm{k} ]-d_{mm}(\omega)\mathcal{A}_{mn}^{\alpha_1}
d_{nm}(\omega_2)  \mathcal{A}_{  nm}^{\alpha_2}f_{  mn}\partial^\alpha \varepsilon_{  mn}.
\end{eqnarray}
Since $\varrho^{1,\alpha_1\alpha_2}$ is diagonal, $\chi_1^{\alpha\alpha_1\alpha_2}$ comes from the combination of the diagonal part of $v^\alpha$ and $\varrho^{1,\alpha_1\alpha_2}$. Owing to the presence of $f_{nm}$, $\chi_2^{\alpha\alpha_1\alpha_2}$($\chi_3^{\alpha\alpha_1\alpha_2}$) comes from  the nondiagonal part of $v$ and $\varrho^{2,\alpha_1\alpha_2}$($\varrho^{3,\alpha_1\alpha_2}$). For $\varrho^{4,\alpha_1\alpha_2}$, the absence of nondiagonal constrain of $f_{nm}$ results the nondiagonal term $\chi_4^{\alpha\alpha_1\alpha_2}$ and the diagonal term $\chi_5^{\alpha\alpha_1\alpha_2}$.

When we consider the static limit, $\chi^{\alpha\alpha_1\alpha_2}_{1}$ is directly obtained as
\begin{equation}
\chi^{\alpha\alpha_1\alpha_2}_{1}\xrightarrow[DC]{}  -\frac{\tau^2e^3}{\hbar^3}\sum_m \int [d \bm{k} ](\partial^{\alpha}\varepsilon_{  m})  \partial^{\alpha_1}\partial^{\alpha_2}f_{  m}.
\end{equation}
One observes that it is found as the Drude contribution with the conductivity tensor $\sigma_{\text{Drude}}^{\alpha\alpha_1\alpha_2}=e^3\chi^{\alpha\alpha_1\alpha_2}_{1}$, which is the dominant contribution to the 2nd-order conductivity in clean metals. This term is finite only if when both of the $\mathcal{P}$ and $\mathcal{T}$ are violated.

For $\chi^{\alpha\alpha_1\alpha_2}_{2}$, whether the $\mathcal{T}$ exists or not affects its dependence on the relaxation time $\tau$.
For scalar particles, $\mathcal{T}$-symmetry renders:
\begin{equation}
\mathcal{A}_{  \bm{k}mn}^{\alpha}=\mathcal{A}_{-\bm{k}  nm}^{\alpha},
\end{equation}
and we obtain
\begin{equation}\label{chi-2-mid-T}
\begin{aligned}
&\chi^{\alpha\alpha_1\alpha_2}_{2}(\omega_1,\omega_2)\xrightarrow[]{\mathcal{T}} \sum_{m,n} \frac{e^3}{\hbar}\int [d \bm{k} ]-\frac{1}{2}[d_{mn}(\omega)-d_{nm}(\omega)] d_{mm}(\omega_2)\varepsilon_{  mn}\mathcal{A}_{  nm}^{\alpha}\mathcal{A}_{  mn}^{\alpha_1}\partial^{\alpha_2}f_{  nm},\\
&\chi^{\alpha\alpha_1\alpha_2}_{5}(\omega_1,\omega_2)\xrightarrow[]{\mathcal{T}} \sum_{m,n} \frac{e^3}{\hbar}\int [d \bm{k} ]-\frac{1}{2}[d_{nm}(\omega_2)-d_{mn}(\omega_2)] d_{mm}(\omega)f_{  mn}\mathcal{A}_{  nm}^{\alpha_2}\mathcal{A}_{  mn}^{\alpha_1}\partial^{\alpha}\varepsilon_{  mn}.
\end{aligned}
\end{equation}
For particles with half-integer spins,  every band is two fold degenerate at each $\bm{k}$ point owing to $\mathcal{T}$, and the inter-band Berry connection are transformed as \cite{PhysRevResearch.2.043081}
\begin{equation}\label{a-t-spinfull}
\mathcal{A}_{M\rho,N\rho^\prime}(\bm{k})=\sum_{\tau,\tau^\prime}\mathcal{A}_{N\tau,M\tau^\prime}(-\bm{k})(-i\sigma_y)^{\dagger}_{\rho^\prime\tau}(-i\sigma_y)_{\tau^\prime\rho}.
\end{equation}
In which the Kramers doublet is explicitly written as $\ket{u_{\bm{k}m}}=\ket{u_{\bm{k}M,\rho}}$ with Kramers degrees of freedom  $\rho=\pm$.
By use of \Eq{a-t-spinfull}, one can verify that  \Eq{chi-2-mid-T} applies to the spinors and thus applies to the particles with arbitrary spin numbers. Within the static limit, we have
\begin{equation}
\frac{1}{2}[d_{mn}(\omega)-d_{nm}(\omega)]d_{mm}(\omega_2)\xrightarrow[\text{DC}]{} i\frac{1}{\varepsilon_{mn}\gamma},
\end{equation}
and $\chi^{\alpha\alpha_1\alpha_2}_{2}$ takes the form
\begin{equation}
\begin{aligned}
\chi^{\alpha\alpha_1\alpha_2}_{2} \xrightarrow[\text{DC}]{\mathcal{T}} &\sum_{m,n} -\frac{\tau e^3}{\hbar}i\int [d\bm{k}  ](\partial^{\alpha_1}f_{  nm})\mathcal{A}_{  mn}^{\alpha}\mathcal{A}_{  mn}^{\alpha_2}\\
=&\sum_{m}\frac{\tau e^3}{\hbar}\int [d\bm{k}  ]- \varepsilon^{\alpha\alpha_2 \gamma
}f_{  m}\partial^{\alpha_1}\Omega_{  m}^{\gamma}.
\end{aligned}
\end{equation}
It is seen that the integral of $\chi^{\alpha\alpha_1\alpha_2}_{2}|_{\omega_1,\omega_2 \rightarrow 0}$ is the Berry curvature dipole $\mathcal{D}^{\alpha_1\gamma}=\int [d \bm{k} ]\varepsilon^{\alpha\alpha_2 \gamma
}f_{  m}\partial^{\alpha_1}\Omega_{  m}^{\gamma}$ \cite{PhysRevLett.115.216806, PhysRevLett.105.026805},  which is also called the extrinsic non-linear Hall effect, since it depends on the relaxation time $\tau$.

For $\chi_5^{\alpha\alpha_1\alpha_2}$ it becomes
\begin{equation}
\begin{aligned}
\chi^{\alpha\alpha_1\alpha_2}_{5}(\omega_1,\omega_2)\xrightarrow[\text{DC}]{\mathcal{T}} \sum_{m,n}\frac{e^3}{\hbar} \int [d \bm{k} ]\frac{\tau}{ \varepsilon_{mn}} f_{  mn}\mathcal{A}_{  nm}^{\alpha_1}\mathcal{A}_{  mn}^{\alpha_2}\partial^{\alpha}\varepsilon_{  mn}.
\end{aligned}
\end{equation}
Noting that $\chi_{5}^{\alpha\alpha_1\alpha_2}$ does not contribute to the conductivity since it vanishes under the symmetrize procedure $[\alpha_1, \omega_1 ] \leftrightarrow [\alpha_2,\omega_2]$.

However, the broken of time-reversal symmetry but maintaining a combined operation of $\mathcal{T}$ and a spacial symmetry may turn $\chi^{\alpha\alpha_1\alpha_2}_{2}$ and $\chi^{\alpha\alpha_1\alpha_2}_{5}$ intrinsic . For instance, let us consider the  $\mathcal{PT}$ symmetry. For the scalar particles, the $\mathcal{PT}$ symmetry yields:
\begin{equation}\label{bc-pt}
\mathcal{A}_{  \bm{k}mn}^{\alpha}=\mathcal{A}_{\bm{k}  nm}^{\alpha}.
\end{equation}
By use of \Eq{bc-pt}, we obtain
\begin{equation}\label{chi-2-mid-PT}
\begin{aligned}
&\chi^{\alpha\alpha_1\alpha_2}_{2}(\omega_1,\omega_2)\xrightarrow[]{\mathcal{PT}}& -\sum_{m,n} \frac{e^3}{\hbar}\int [d \bm{k} ]\frac{1}{2}(d_{mn}(\omega)+d_{nm}(\omega)) d_{mm}(\omega_2)\varepsilon_{  mn}\mathcal{A}_{  nm}^{\alpha}\mathcal{A}_{  mn}^{\alpha_1}\partial^{\alpha_2}f_{  nm},\\
&\chi^{\alpha\alpha_1\alpha_2}_{5}(\omega_1,\omega_2)\xrightarrow[]{\mathcal{PT}}& -\sum_{m,n} \frac{e^3}{\hbar}\int [d \bm{k} ]\frac{1}{2}[d_{nm}(\omega_2)+d_{mn}(\omega_2)] d_{mm}(\omega)f_{  mn}\mathcal{A}_{  nm}^{\alpha_1}\mathcal{A}_{  mn}^{\alpha_2}\partial^{\alpha}\varepsilon_{  mn}.
\end{aligned}
\end{equation}
Taking the static limit, we have
\begin{equation}
\begin{aligned}
&\frac{1}{2}[d_{mn}(\omega)+d_{nm}(\omega)]d_{mm}(\omega_2)\xrightarrow[\text{DC}]{} \frac{1}{2(\omega_2 + i\gamma)}\frac{2(\omega+2i\gamma)}{(\omega+2i\gamma)^2 -\varepsilon_{nm}^2}\rightarrow -\frac{2}{\varepsilon_{nm}^2},\\
&\frac{1}{2}[d_{mn}(\omega_2)+d_{nm}(\omega_2)]d_{mm}(\omega)\xrightarrow[\text{DC}]{} \frac{1}{2(\omega + i\gamma)}\frac{2(\omega_2+i\gamma)}{(\omega_2+i\gamma)^2 -\varepsilon_{nm}^2}\rightarrow -\frac{1}{2\varepsilon_{nm}^2}.
\end{aligned}
\end{equation}
In above, the clean limit ($\gamma \rightarrow 0$) is take since the expressions converge. 
For the spinors, the $\mathcal{PT}$ symmetry results
\begin{equation}
\mathcal{A}_{M\rho,N\rho^\prime}(\bm{k})=-\sum_{\tau,\tau^\prime}\mathcal{A}_{N\tau,M\tau^\prime}(\bm{k})(-i\sigma_y)^{\dagger}_{\rho^\prime\tau}(-i\sigma_y)_{\tau^\prime\rho}.
\end{equation}
As expected, it is found that \Eq{chi-2-mid-PT} applies to particles with integer and half-integer spins. Following the parallel demonstration, we obtain
\begin{equation}\label{pi3-pt}
\begin{aligned}
&\chi^{\alpha\alpha_1\alpha_2}_{2} \xrightarrow[\text{DC}]{\mathcal{PT}}\sum_{m,n} \frac{e^3}{\hbar}  \int [d\bm{k}]\frac{2}{\varepsilon_{mn  }}\mathcal{A}_{  nm}^{\alpha}\mathcal{A}_{  mn}^{\alpha_1}\partial^{\alpha_2}f_{  nm}=\sum_{m,n} - \frac{e^3}{\hbar} \int [d\bm{k}]\frac{4}{\varepsilon_{mn}}g^{\alpha\alpha_1}_{mn} \partial^{\alpha_2} f_{  m},\\
&\chi^{\alpha\alpha_1\alpha_2}_{5} \xrightarrow[\text{DC}]{\mathcal{PT}}\sum_{m,n}  \frac{e^3}{\hbar}\frac{1}{2\varepsilon_{mn}^2}f_{mn  } \mathcal{A}_{  nm}^{\alpha_1}\mathcal{A}_{  mn}^{\alpha_2}\partial^{\alpha}\varepsilon_{mn}=-\sum_{m,n}  \frac{e^3}{\hbar}\partial^{\alpha}(\frac{1}{\varepsilon_{mn}}) g^{\alpha_1\alpha_2}_{mn}f_{m}.
\end{aligned}
\end{equation}
In which the quantum metric is defined as $ g^{\alpha_1\alpha_2}_{mn}=\frac{1}{2}(\mathcal{A}_{mn}^{\alpha_1}\mathcal{A}_{nm}^{\alpha_2}+\mathcal{A}_{mn}^{\alpha_2}\mathcal{A}_{nm}^{\alpha_1})$.
which manifests itself as the intrinsic contribution to the 2nd-order conductivity tensor since it is independent on the relaxation time $\tau$. As mentioned, consider $\chi^{\alpha\alpha_1\alpha_2}_{2}$ is not symmetric, and it should be symmetrized with respect to the external field indices $\alpha_1$ and $\alpha_2$ as
\begin{equation}
\chi^{\alpha\alpha_1\alpha_2}_{2} =\sum_{m,n} - \frac{e^3}{\hbar} \int [d\bm{k}]\frac{2}{\varepsilon_{mn}}(g^{\alpha\alpha_1}_{mn} \partial^{\alpha_2} + g^{\alpha\alpha_2}_{mn} \partial^{\alpha_1})f_{  m},
\end{equation}

Next we give the symmetry analysis of $\chi^{\alpha\alpha_1\alpha_2}_{3}$. Since this term is not divergent in the clean limit, the formula in static limit can be written as:
\begin{equation}
\chi^{\alpha\alpha_1\alpha_2}_{3}\xrightarrow[\text{DC}]{}  \sum_{m,n}-\frac{e^3}{\hbar}\int [d \bm{k} ]\partial^{\alpha_1}\left(\frac{1}{\varepsilon_{  mn}}  \mathcal{A}_{  mn}^{\alpha_2}f_{  nm}\right)\mathcal{A}_{  nm}^{\alpha},
\end{equation}
The $\mathcal{T}$ symmetry results:
\begin{equation}
 \partial^{\alpha_1}\left(\frac{ \mathcal{A}_{ \bm{k} mn}^{\alpha_2}f_{\bm{k}  mn}}{\varepsilon_{\bm{k}  mn}} \right)\mathcal{A}_{\bm{k}  nm}^{\alpha}=-\partial^{\alpha_1}\left(\frac{\mathcal{A}_{- \bm{k} nm}^{\alpha_2}f_{\bm{-k}mn}}{\varepsilon_{-  \bm{k}mn}}  \right)\mathcal{A}_{-\bm{k}  mn}^{\alpha}.
\end{equation}
By using the obtained relations, one can verify that $\chi^{\alpha\alpha_1\alpha_2}_{ 3}$ vanishes under $\mathcal{T}$. For the $\mathcal{PT}$-symmetric system, the $\mathcal{PT}$ symmetry gives:
\begin{equation}\label{pt-pi2}
 \partial^{\alpha_1}\left(\frac{ \mathcal{A}_{\bm{k}  mn}^{\alpha_2}f_{ \bm{k} mn}}{\varepsilon_{  mn}} \right)\mathcal{A}_{ \bm{k} nm}^{\alpha}=\partial^{\alpha_1}\left(\frac{\mathcal{A}_{ \bm{k} nm}^{\alpha_2}f_{ \bm{k} mn}}{\varepsilon_{  mn}}  \right)\mathcal{A}_{ \bm{k} mn}^{\alpha}.
\end{equation}
Making use of \Eq{pt-pi2}, we obtain:
\begin{equation}\label{pt-pi2-r}
\chi^{\alpha\alpha_1\alpha_2}_{3}\xrightarrow[\text{DC}]{\mathcal{PT}}  \sum_{m,n}\frac{e^3}{\hbar}\int [d \bm{k} ]-\frac{1}{2} f_{  mn}\left(\frac{\mathcal{A}_{  mn}^{\alpha_2}\mathcal{B}_{  nm}^{\alpha_1 \alpha}+\mathcal{B}_{  mn}^{\alpha_1 \alpha}\mathcal{A}_{  nm}^{\alpha_2}}{\varepsilon_{  mn}}\right),
\end{equation}
where the dipole of the inter-band Berry connection is defined as $\mathcal{B}_{  mn}^{\alpha\beta}=\partial^{\alpha}\mathcal{A}_{  mn}^{\beta}$. By use of the relation
\begin{equation}
\partial^\alpha \mathcal{A}_{mn}^\beta - \partial^\beta \mathcal{A}_{mn}^\alpha = i\langle \partial^\alpha u_m |\partial^\beta u_n\rangle - i\langle \partial^\beta u_m |\partial^\alpha u_n\rangle ,
\end{equation}
\Eq{pt-pi2-r} is rewritten as
\begin{equation}
\chi^{\alpha\alpha_1\alpha_2}_{3}=\pi^{\alpha\alpha_1\alpha_2} + \eta^{\alpha\alpha_1\alpha_2}
\end{equation}
in which
\begin{equation}
\begin{aligned}
&\pi^{\alpha\alpha_1\alpha_2} = \sum_{m,n}-\frac{e^3}{\hbar}\int [d \bm{k} ]f_{  mn}\left(\frac{\mathcal{A}_{  mn}^{\alpha_2}\mathcal{B}_{  nm}^{\alpha \alpha_1}}{\varepsilon_{  mn}}\right),\\
&\eta^{\alpha\alpha_1\alpha_2} = \sum_{m,n,p}-\frac{e^3}{\hbar}i\int [d \bm{k} ]f_{  mn}\mathcal{A}_{  mn}^{\alpha_2}\left(\frac{\mathcal{A}_{np}^{\alpha_1}\mathcal{A}_{pm}^{\alpha}-\mathcal{A}_{np}^{\alpha}\mathcal{A}_{pm}^{\alpha_1}}{\varepsilon_{  mn}}\right),
\end{aligned}
\end{equation}
where we use
\begin{equation}
\langle \partial^{\alpha_1} u_n|\partial^{\alpha}u_m\rangle - \langle \partial^{\alpha} u_n|\partial^{\alpha_1}u_m\rangle = \sum_p \mathcal{A}_{np}^{\alpha_1}\mathcal{A}_{pm}^{\alpha}-\mathcal{A}_{np}^{\alpha}\mathcal{A}_{pm}^{\alpha_1}.
\end{equation}
By using integration by parts, we can separate $\pi^{\alpha\alpha_1\alpha_2}$ into two parts with $\pi^{\alpha\alpha_1\alpha_2} = \pi_{1}^{\alpha\alpha_1\alpha_2}+\pi_{2}^{\alpha\alpha_1\alpha_2}$:
\begin{eqnarray}
\pi_{1}^{\alpha\alpha_1\alpha_2} &=& \sum_{m,n}\frac{e^3}{\hbar}\int [d \bm{k} ]\partial^\alpha f_{  mn}\left(\frac{\mathcal{A}_{  mn}^{\alpha_2}\mathcal{A}_{  nm}^{\alpha_1}}{\varepsilon_{  mn}}\right), \label{pi-1}\\
\pi_{2}^{\alpha\alpha_1\alpha_2} &=& \sum_{m,n}\frac{e^3}{\hbar}\int [d \bm{k} ] f_{  mn}\partial^\alpha\left(\frac{\mathcal{A}_{  mn}^{\alpha_2}}{\varepsilon_{  mn}}\right)\mathcal{A}_{  nm}^{\alpha_1}. \label{pi-2}
\end{eqnarray}
Combining \Eq{pi-1} and \Eq{pi-2}, we have 
\begin{equation}
\pi^{\alpha\alpha_1\alpha_2} = -\sum_{m,n}\frac{e^3}{\hbar}\int [d \bm{k} ]\frac{1}{\varepsilon_{mn}}\partial^\alpha g_{mn}^{\alpha_1\alpha_2} f_{  m}, 
\end{equation}
Now we write the intrinsic 2nd-order conductivity tensor  as
\begin{equation}
\sigma^{\alpha\alpha_1\alpha_2}_{\text{int}}=\chi^{\alpha\alpha_1\alpha_2}_2 + \pi^{\alpha\alpha_1\alpha_2}  + \eta^{\alpha\alpha_1\alpha_2} + \chi_{4}^{\alpha\alpha_1\alpha_2}+\chi_{5}^{\alpha\alpha_1\alpha_2}.
\end{equation}
$\chi^{\alpha\alpha_1\alpha_2}_{4}(\omega_1,\omega_2)$ in the static limit is
\begin{equation}\label{pi2-4}
\begin{aligned}
\chi^{\alpha\alpha_1\alpha_2}_{4  }\xrightarrow[\text{DC}]{}
 \sum_{m,n,p} \frac{e^3}{\hbar}\int [d \bm{k} ]i\mathcal{A}_{  nm}^{\alpha} \left(\mathcal{A}_{  mp}^{\alpha_1}\frac{1}{\varepsilon_{  pn}}\mathcal{A}_{  pn}^{\alpha_2}f_{  np} -\mathcal{A}_{  pn}^{\alpha_1}\frac{1}{\varepsilon_{  mp}}\mathcal{A}_{  mp}^{\alpha_2}f_{  pm}\right).
\end{aligned}
\end{equation}
Following a parallel discussion, one found $\chi_{4}^{\alpha\alpha_1\alpha_2}$ vanishes under $\mathcal{T}$. For the $\mathcal{PT}$ symmetry, one can verify (by use of band-index permutation), we have:
\begin{equation}
\begin{aligned}
& \chi_{4}^{\alpha\alpha_1\alpha_2}+\eta^{\alpha\alpha_1\alpha_2}\\
=&\sum_{m,n,p}\frac{e^3}{\hbar}\int [d \bm{k} ]if_{  pn}\mathcal{A}_{  pn}^{\alpha_2}\frac{\mathcal{A}_{  nm}^{\alpha_1}\mathcal{A}_{  mp}^{\alpha}+ \mathcal{A}_{  nm}^{\alpha}\mathcal{A}_{  mp}^{\alpha_1}}{\varepsilon_{  pn}}-\sum_{m,n,p}\frac{e^3}{\hbar}\int [d \bm{k} ]if_{  mp}\mathcal{A}_{  mp}^{\alpha_2}\frac{\mathcal{A}_{  pn}^{\alpha_1}\mathcal{A}_{  nm}^{\alpha}+ \mathcal{A}_{  pn}^{\alpha}\mathcal{A}_{  nm}^{\alpha_1}}{\varepsilon_{  mp}}\\
=&0.
\end{aligned}
\end{equation}
It is found that with $\mathcal{PT}$ symmetry, the term involving three-band transitions cancels out, and the 2nd-order conductivity is composed of the terms up to two band transitions. 
Now, we have
\begin{equation}
\sigma^{\alpha\alpha_1\alpha_2}_{\text{int}}=\chi^{\alpha\alpha_1\alpha_2}_2 + \pi^{\alpha\alpha_1\alpha_2} +\chi_{5}^{\alpha\alpha_1\alpha_2},
\end{equation}
which can be reorganized as
\begin{equation}\label{re-org}
\sigma^{\alpha\alpha_1\alpha_2}_{\text{int}}=(\frac{1}{2}\chi^{\alpha\alpha_1\alpha_2}_2 + 2\pi^{\alpha\alpha_1\alpha_2} +2\chi_{5}^{\alpha\alpha_1\alpha_2}) + (\frac{1}{2}\chi^{\alpha\alpha_1\alpha_2}_2 - \pi^{\alpha\alpha_1\alpha_2} -\chi_{5}^{\alpha\alpha_1\alpha_2}) ,
\end{equation}
In which the first term is contribution of the anomalous nonlinear Hall effect (or, alternatively referred as the intrinsic nonlinear Hall effect):
\begin{equation}
\begin{aligned}
\sigma^{\alpha\alpha_1\alpha_2}_{\rm{INAHE}} =
& \frac{1}{2}\chi^{\alpha\alpha_1\alpha_2}_2 + 2\pi^{\alpha\alpha_1\alpha_2} +2\chi_{5}^{\alpha\alpha_1\alpha_2}\\
=&  \sum_{m}\frac{e^3}{\hbar}\int [d \bm{k} ](- \partial^\alpha  G_{m}^{\alpha_1\alpha_2}f_{  m} + \partial^{\alpha_2} G_{m}^{\alpha\alpha_1} f_{  m}) + (\alpha_1 \leftrightarrow \alpha_2)\\
\end{aligned}
\end{equation}
where the Berry-connection polarizibility is defined as \cite{PhysRevLett.127.277202}
\begin{equation}\label{bcp-def}
G_{m}^{\alpha_1\alpha_2}=\sum_{n}2\frac{\mathcal{A}_{  mn}^{\alpha_1}\mathcal{A}_{  nm}^{\alpha_2}+\mathcal{A}_{  mn}^{\alpha_2}\mathcal{A}_{  nm}^{\alpha_1}}{\varepsilon_{  mn}}=\sum_{n}2\text{Re}\frac{\mathcal{A}_{  mn}^{\alpha_1}\mathcal{A}_{  nm}^{\alpha_2}}{\varepsilon_{  mn}}.
\end{equation}
It manifests itself as the intrinsic contribution to the 2nd-order Hall conductivity \cite{PhysRevLett.112.166601, PhysRevLett.127.277201}. The second term in \Eq{re-org} is the intrinsic nonlinear Ohmic current, which is recognised as
\begin{equation}
\begin{aligned}
\sigma^{\alpha\alpha_1\alpha_2}_{\rm{INOC}} =
& \frac{1}{2}\chi^{\alpha\alpha_1\alpha_2}_2 - \pi^{\alpha\alpha_1\alpha_2} -\chi_{5}^{\alpha\alpha_1\alpha_2}\\
=&  \sum_{m}\frac{e^3}{\hbar}\int [d \bm{k} ](\partial^\alpha G_{m}^{\alpha_1\alpha_2} + \partial^{\alpha_1} G_{m}^{\alpha\alpha_2} +\partial^{\alpha_2} G_{m}^{\alpha\alpha_1})f_{  m}.\\
\end{aligned}
\end{equation}

Table \ref{2ndct} shows the classification results of the 2nd-order electric responses in the $\mathcal{T}$-/$\mathcal{PT}$-symmetry systems.
\begin{table*}[hptb]
	\renewcommand\arraystretch{2}
	\caption{ Classification of the  2nd-order DC electric conductivity in the $\mathcal{T}$-/$\mathcal{PT}$-symmetry systems.}\label{sigma-t-pt}
	\begin{tabular*}{10cm}{@{\extracolsep{\fill}}p{0.8cm}cccc}
		\hline\hline
      & $\mathcal{T}$ & $\mathcal{PT}$ &
      \\		\hline
      $\chi_1$ &$\times$ & $\sigma_{\text{Drude}}$
      \\
	  $\chi_2$ & $\sigma_{\text{BCD}}$ &$\chi_2 (\mathcal{PT})$  & \multirow{2}{*}{$\frac{1}{2}\chi_2(\mathcal{PT}) + 2\pi +2\chi_{5}(\mathcal{PT})=\sigma_{\text{INAHE}}$}
	  \\	
	  $\chi_3$ & $\times$ & $\pi  + \eta$ & \multirow{2}{*}{$\frac{1}{2}\chi_2(\mathcal{PT}) - \pi +\chi_{5}(\mathcal{PT})=\sigma_{\text{INOC}}$}
	  \\	
	  $\chi_4$ & $\times$ & $\chi_4 (\mathcal{PT})$ & \multirow{2}{*}{$\eta  + \chi_4 (\mathcal{PT})=0$}
      \\
      $\chi_5$ & $\times$  & $\chi_5(\mathcal{PT})$\\
		\hline\hline
	\end{tabular*}\label{2ndct}
\end{table*}

\section{Symmetry transformations of the four terms of the 2nd-order conductivity}\label{sym-ana}
For the 2nd-order Drude conductivity tensor, it is $\mathcal{T}$-odd and manifests as the time-pseudotensor. The crystal symmetries impose constraints of the form
\begin{equation}\label{drude-st}
\sigma_{\text{Drude}}^{\alpha\beta\gamma} =  \sum_{\delta\nu\xi}D^{\alpha\delta}D^{\beta\mu}D^{\gamma\xi}\sigma_{\text{Drude}}^{\delta\nu\xi},
\end{equation}
In addition, one finds that $\sigma_{\text{Drude}}^{\alpha\beta\gamma}$ is symmetric with in all space indices, as a result only 18 of the $3^3$ elements are independent.
The BCD is $\mathcal{T}$-odd and is an antisymmetric tensor
\begin{equation}
\Sigma_{\text{BCD}}^{\gamma\delta}=\frac{1}{2}\epsilon^{\alpha\beta\gamma}\sigma_{\text{BCD}}^{\alpha\beta\delta}.
\end{equation}
The constraints of the crystal symmetries take the form:
\begin{equation}\label{bcd-st}
\Sigma_{\text{BCD}}^{\alpha\beta} = {\rm{det}}(D) \sum_{\gamma\delta}D^{\alpha\gamma}D^{\beta\delta}\Sigma_{\text{BCD}}^{\gamma\delta}.
\end{equation}
\Eq{bcd-st} indicates that $\Sigma_{\text{BCD}}$ behaves like the space-pseudotensor.
For $\sigma_{\text{INAHE}}$, due to the fact that it is antisymmetric by permuting the first two indices
\begin{equation}
\Sigma_{\text{INAHE}}^{\gamma\delta}=\frac{1}{2}\epsilon^{\alpha\beta\gamma}\sigma_{\text{INAHE}}^{\alpha\beta\delta},
\end{equation}
And the transformations under the point group operations take the form
\begin{equation}\label{bcp-st}
\Sigma_{\text{INAHE}}^{\alpha\beta} = \eta_{T}{\rm{det}}(D) \sum_{\gamma\delta}D^{\alpha\gamma}D^{\beta\delta}\sigma_{\text{INAHE}}^{\gamma\delta},
\end{equation}
therefore it is recognized as the time-space-pseudotensor. By use of symmetry transformations of $\sigma_{\text{INOC}}$ given in Eq. (4) in the main text, the constrains of the generators of MPGs on the INOC conductivity are obtained ( Table I  in the main text). Combined with \Eq{drude-st}, \Eq{bcd-st} and \Eq{bcp-st}, the constrains of the MPGs on the all four terms of 2nd order electric conductivity tensors are obtained, as listed in Table \ref{table2}. It can be seen that the part excluding $\sigma_{\text{INOC}}$ in Table \ref{table2} is just the case discussed in Ref.\cite{PhysRevLett.127.277201}.

\begin{table}[h]
	\renewcommand\arraystretch{2}
	\caption{Magnetic point groups classied by the existence or absence of the conductivities of the nonlinear Drude current , BCD induced nonlinear current, intrinsic nonlinear anomalous Hall current, and the INOC contribution.}
	\begin{tabular*}{18cm}{@{\extracolsep{\fill}}p{10cm}cccc}
		\hline\hline
         MPGs & $\sigma_{\text{INAHE}}$ & $\sigma_{\text{Drude}}$ & $\sigma_{\text{BCD}}$&$\sigma_{\text{INOC}}$
          \\
		\hline
        $4/\text{m}^{\prime} \text{m}^{\prime} \text{m}^{\prime} , \bar{6}^{\prime} \text{m}^{\prime} 2, 6/\text{m}^{\prime} \text{m}^{\prime} \text{m}^{\prime} $
        &$\checkmark$ & $\times$ &$\times$ &$\checkmark$\\
        \hline
	  $\bar{6},6^{\prime} /\text{m},\bar{6}\text{m}2,\bar{6}\text{m}^{\prime} 2^{\prime} ,6^{\prime} /\text{m}\text{m}\text{m}^{\prime} ,23,\text{m}^{\prime} \bar{3},4^{\prime} 32^{\prime} ,\bar{4}3\text{m},\text{m}^{\prime} \bar{3}\text{m}$
	  & $\times$ &$\checkmark$  &$\times$ &$\checkmark$\\
	  \hline
	  $11^{\prime} ,21^{\prime} ,31^{\prime} ,41^{\prime} ,61^{\prime} ,\text{m}1^{\prime} ,2\text{m}\text{m}1^{\prime} ,3\text{m}1^{\prime} ,4\text{m}\text{m}1^{\prime} ,$
	  &\multirow{2}*{ $\times$}  &\multirow{2}*{ $\times$} &\multirow{2}*{ $\checkmark$}&\multirow{2}*{ $\times$}  \\
	  $6\text{m}\text{m}1^{\prime} ,2221^{\prime} ,321^{\prime} ,4221^{\prime} ,6221^{\prime} ,\bar{4}2\text{m}1^{\prime} ,\bar{4}1^{\prime} $\\
	  \hline
	  $\bar{1}^{\prime}, 2^{\prime} /\text{m}, 2 /\text{m}^{\prime},\text{m}^{\prime}\text{m}\text{m},\text{m}^{\prime}\text{m}^{\prime}\text{m}^{\prime}, 4 /m^{\prime}$, $4^{\prime} /\text{m}^{\prime}, 4 /\text{m}^{\prime}\text{m}\text{m}, $
 &\multirow{2}*{$\checkmark$} &\multirow{2}*{ $\checkmark$} &\multirow{2}*{ $\times$}&\multirow{2}*{ $\checkmark$} \\
	  $4^{\prime}/\text{m}^{\prime}\text{m}^{\prime}\text{m},\bar{3}^{\prime},\bar{3}^{\prime}\text{m},\bar{3}^{\prime}\text{m}^{\prime} ,\bar{6},6 /\text{m},\bar{6}^{\prime}\text{m} 2,6 /\text{m}^{\prime}\text{m}\text{m}$
	  \\
	  \hline
	  $422, 4 \text{m}^{\prime} \text{m}^{\prime}, \bar{4}^{\prime} 2 \text{m}^{\prime}, 622, 6 \text{m}^{\prime} \text{m}^{\prime}$
	  &$\checkmark$ & $\times$ &$\checkmark$ & $\checkmark$\\
	  \hline
	  $6^{\prime}, 6^{\prime} 22^{\prime}, 6^{\prime} \mathrm{mm}^{\prime}$
	 & $\times$   &$\checkmark$ &$\checkmark$&$\checkmark$\\
	  \hline
	  $1,2,2^{\prime}, \mathrm{m}, \mathrm{m}^{\prime}, 222,2^{\prime} 2^{\prime} 2, \mathrm{mm} 2, \mathrm{m}^{\prime} \mathrm{m} 2^{\prime},\mathrm{m}^{\prime} \mathrm{m}^{\prime} 2,$
	 &\multirow{3}*{$\checkmark$} &\multirow{3}*{ $\checkmark$} &\multirow{3}*{ $\checkmark$} &\multirow{3}*{ $\checkmark$} \\
	  $4,4^{\prime},\bar{4},\bar{4}^{\prime}, 4^{\prime} 22^{\prime}, 42^{\prime} 2^{\prime},4 \mathrm{mm}, 4^{\prime} \mathrm{m}^{\prime} \mathrm{m}, \bar{4}2 \mathrm{m},\bar{4}^{\prime} 2^{\prime} \mathrm{m},$\\
	  $\bar{4}2^{\prime} \mathrm{m},3,32,32^{\prime}, 3\mathrm{m}, 3 \mathrm{m}^{\prime}, 6, 62^{\prime} 2^{\prime}, 6 \mathrm{mm}$\\
		\hline\hline
	\end{tabular*}
	\label{table2}
\end{table}

\bibliographystyle{apsrev4-2}
%

\end{widetext}

\end{document}